\pdfoutput=1

\documentclass[sigconf]{acmart}

\AtBeginDocument{%
  \providecommand\BibTeX{{%
    \normalfont B\kern-0.5em{\scshape i\kern-0.25em b}\kern-0.8em\TeX}}}

\copyrightyear{2024}
\acmYear{2024}
\acmConference[LLM4Eval@SIGIR2024]{The First Workshop on Large Language Model for Evaluation in Information Retrieval}{July 18, 2024}{Washington D.C., USA}
\acmBooktitle{The First Workshop on Large Language Model for Evaluation in Information Retrieval (LLM4Eval@SIGIR2024), July 18, 2024, Washington D.C., USA}
\acmDOI{}
\acmPrice{}
\acmISBN{}
\setcopyright{rightsretained}

\usepackage{microtype}
\usepackage{hyperref}
\usepackage{url}
\usepackage{booktabs}

\usepackage{times}
\usepackage{latexsym}

\usepackage[T1]{fontenc}
\usepackage[utf8]{inputenc}

\usepackage{enumerate}
\usepackage{graphicx}
\usepackage{subcaption}
\usepackage{graphics, amsmath}
\usepackage{multirow}

\usepackage{tabularx}
\usepackage{pifont}
\usepackage{colortbl}

\usepackage{xcolor}
\usepackage[inline]{enumitem}
\usepackage{makecell}
\usepackage{adjustbox}

\usepackage{acronym}

\usepackage{siunitx}
\usepackage{array}
\usepackage{caption}
\usepackage{wrapfig}
\usepackage{xspace}

\usepackage{soul}

\newcommand{\gense}{VeJudge\xspace}

\newcommand{\factcc}{\textsc{FactCC}\xspace}
\newcommand{\bertscore}{\textsc{BERTScore}\xspace}
\newcommand{\bartscore}{\textsc{BARTScore}\xspace}
\newcommand{\summac}{\textsc{SummaC}\xspace} %
\newcommand{\summacconvfull}{\textsc{SummaC}\textsubscript{Conv}\xspace}
\newcommand{\summaczsfull}{\textsc{SummaC}\textsubscript{ZS}\xspace}

\newcommand{\autoais}{\textsc{AutoAIS}\xspace}
\newcommand{\alignscore}{\textsc{AlignScore}\xspace}

\newcommand{\isfullno}{\textsc{FS-vs-NS}\xspace}
\newcommand{\ispartialfull}{\textsc{FS-vs-PS}\xspace}
\newcommand{\ispartialno}{\textsc{PS-vs-NS}\xspace}

\begin{document}

\title[A Comparative Analysis of Faithfulness Metrics]{A Comparative Analysis of Faithfulness Metrics and Humans in Citation Evaluation}

\author{Weijia Zhang}
\affiliation{
  \institution{University of Amsterdam}
  \city{Amsterdam}
  \country{Netherlands}
}

\author{Mohammad Aliannejadi}
\affiliation{
  \institution{University of Amsterdam}
  \city{Amsterdam}
  \country{Netherlands}
}

\author{Jiahuan Pei}
\affiliation{%
  \institution{Centrum Wiskunde \& Informatica}
  \city{Amsterdam}
  \country{Netherlands}
}

\author{Yifei Yuan}
\affiliation{
  \institution{University of Copenhagen}
  \city{Copenhagen}
  \country{Denmark}
}

\author{Jia-Hong Huang}
\affiliation{
  \institution{University of Amsterdam}
  \city{Amsterdam}
  \country{Netherlands}
}

\author{Evangelos Kanoulas}
\affiliation{
  \institution{University of Amsterdam}
  \city{Amsterdam}
  \country{Netherlands}
}

\renewcommand{\shortauthors}{Zhang, et al.}

\begin{abstract}

Large language models (LLMs) often generate content with unsupported or unverifiable content, known as ``hallucinations.''
To address this, retrieval-augmented LLMs are employed to include citations in their content, grounding the content in verifiable sources.
Despite such developments, manually assessing how well a citation supports the associated statement remains a major challenge.
Previous studies tackle this challenge by leveraging faithfulness metrics to estimate citation support automatically. 
However, they limit this citation support estimation to a binary classification scenario, neglecting fine-grained citation support in practical scenarios.
To investigate the effectiveness of faithfulness metrics in fine-grained scenarios, we propose a comparative evaluation framework that assesses the metric effectiveness in distinguishing citations between three-category support levels: \textit{full}, \textit{partial}, and \textit{no} support.
Our framework employs correlation analysis, classification evaluation, and retrieval evaluation to measure the alignment between metric scores and human judgments comprehensively.
Our results indicate no single metric consistently excels across all evaluations, highlighting the complexity of accurately evaluating fine-grained support levels. 
Particularly, we find that the best-performing metrics struggle to distinguish partial support from full or no support.
Based on these findings, we provide practical recommendations for developing more effective metrics.

\end{abstract}



\maketitle

\begin{figure}[!b]
    \centering
    \includegraphics[width=\linewidth]{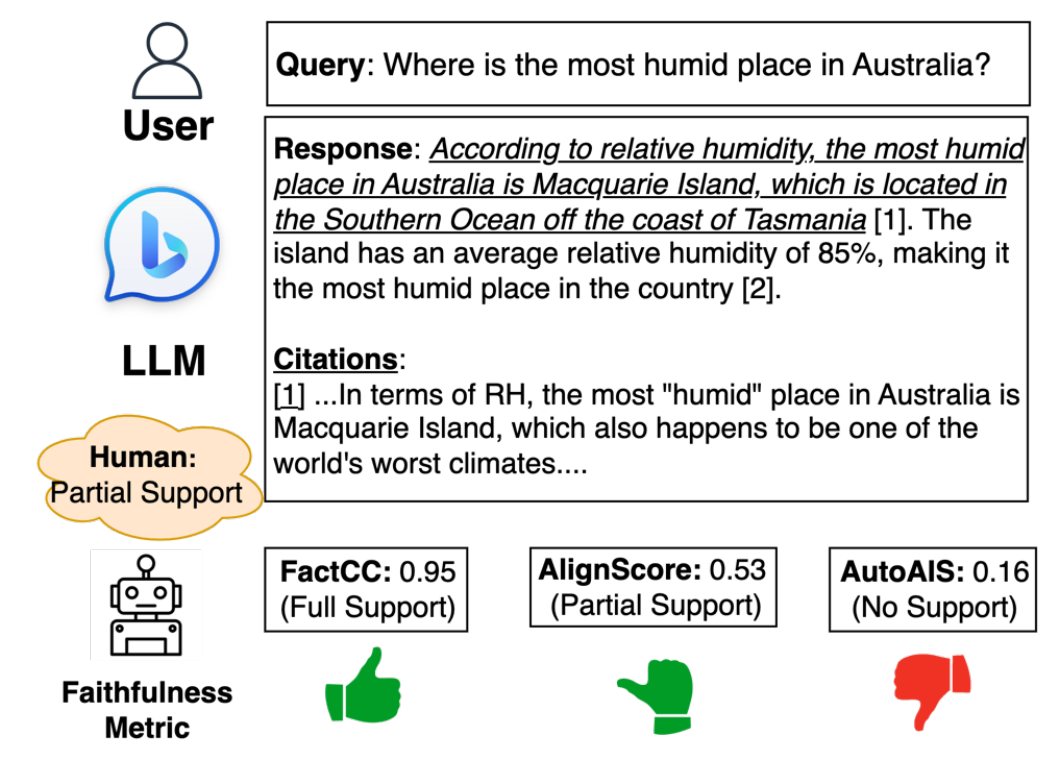}
    \caption{An example of \textit{partial support} in citation evaluation. A retrieval-augmented LLM generates a response that includes citations based on a given user query. The human assessor annotates that the first citation partially supports the associated statement. Inconsistent metric scores are observed when assessing the statement using three distinct faithfulness metrics.}
    \label{fig:example}
\end{figure}

\section{Introduction}

Large language models (LLMs) suffer from generating content known as ``hallucinations''~\cite{zhang2023hallucination}, which refers to content that either contradicts established world knowledge or cannot be verified by any reliable source of information.
Mainstream studies~\cite{bohnet2022attributed,gao-etal-2023-enabling,li2023llatrieval} aims to mitigate this issue by leveraging retrieval-augmented LLMs to generate responses with in-line citations, which contain supporting evidence to verify the statements in the responses.
One primary challenge in this field is to assess how well a citation supports its associated statement, since manually evaluating citations is labor-intensive and time-consuming.
To this end, automated citation evaluation has been explored to minimize reliance on human assessments~\cite{bohnet2022attributed,li2023llatrieval}. 
Given the early stage of this research, faithfulness evaluation metrics~\cite{maynez-etal-2020-faithfulness,kryscinski-etal-2020-evaluating,honovich-etal-2022-true-evaluating} have been employed as proxies to automatically estimate the support levels of the citations~\cite{gao-etal-2023-rarr,gao-etal-2023-enabling}. 
This is motivated by the observation that these metrics measure the extent to how faithful the model-generated text is to the sourced text, aligning closely with the objectives of automated citation evaluation.

Prior studies~\cite{gao-etal-2023-rarr,gao-etal-2023-enabling} have primarily limited the application of faithfulness metrics in automated citation evaluation to a binary classification scenario. 
In this scenario, faithfulness metrics are solely tasked with determining whether a citation supports the associated statement. 
This binary approach fails to capture the fine-grained citation support encountered in real-world applications. 
For instance, consider a ``\textit{partial support}'' scenario illustrated in~\autoref{fig:example}. Given a user query ``\textit{Where is the most humid place in Australia?}'', a retrieval-augmented LLM generates a response along with multiple citations.
A human assessor categorizes the first citation as ``partial support'' since it only supports the initial segment of the statement: ``\textit{the most humid place in Australia is Macquarie Island}''. However, it does not provide evidence for the latter part of the statement: ``\textit{which is located in the Southern Ocean off the coast of Tasmania}''. The complexity of this partial support scenario leads to noticeable inconsistencies across three distinct faithfulness metrics.
However, the effectiveness of faithfulness metrics in accurately distinguishing citations in such fine-grained citation support scenarios remains largely under-explored.

To address the issue above, we propose a comparative evaluation framework designed to assess the effectiveness of faithfulness metrics against human judgments in fine-grained levels of support scenarios. 
In our framework, we define ``\textit{support levels}'' as the extent to which a citation supports the corresponding statement~\cite{liu-etal-2023-evaluating,yue-etal-2023-automatic}. 
More specifically, in contrast to previous studies that predominantly focus on binary classification scenarios, our framework aims to evaluate the effectiveness of faithfulness metrics in a three-category support level scenario: \textit{full support}, \textit{partial support}, and \textit{no support}.
These categories indicate whether a citation provides \textit{full}, \textit{partial}, or \textit{no} support to the associated statement.
To comprehensively assess the metric effectiveness, we measure the alignment between metric scores and human judgments by employing three distinct types of evaluation protocols:
\begin{enumerate*}[label=\arabic*)]
    \item \textit{Correlation analysis:} we employ a standard correlation analysis to determine the extent to which metric scores correlate with human judgments. This analysis highlights the general trend in the relationship between these two variables, offering insights into their alignment;
    \item \textit{Classification evaluation:} we conduct a classification evaluation to assess the metrics' capability to distinguish citations based on their support levels. This evaluation specifically measures the accuracy of the metrics in distinguishing between partial, full, and no support scenarios, providing a clear indication of their effectiveness in three-way classification scenarios; and
    \item \textit{Retrieval evaluation:} we undertake a retrieval evaluation to assess the effectiveness of metrics in ranking citations according to their support levels. This is motivated by the observation that the previous two evaluation protocols assume citations are present within statements. However, this assumption is not always valid in practical applications~\cite{liu-etal-2023-evaluating}. In these cases, faithfulness metrics are adapted to retrieve potential citations from a pool of candidates~\cite{gao-etal-2023-rarr,gao-etal-2023-enabling}. The retrieval evaluation thus plays a pivotal role in determining the practical utility of metric adaptations.
\end{enumerate*}

In our experiments, we assess seven widely used faithfulness metrics, categorizing them into \textit{similarity-based} and \textit{entailment-based} metrics. Our experimental findings are as follows:
\begin{enumerate*}[label=\arabic*)]
    \item no single faithfulness metric consistently outperforms others across three evaluation protocols. This suggests that these protocols are complementary and should be integrated to provide a comprehensive evaluation of metric performance;
    \item the best-performing metrics like the entailment-based \autoais show promising results in distinguishing between full-support and no-support scenarios. Nonetheless, they struggle to identify cases of partial support, highlighting the inherent complexities of automated citation evaluation; and
    \item in terms of retrieval evaluation, similarity-based metrics, such as \bertscore, consistently surpass best-performing entailment-based metrics. This indicates that entailment-based metrics exhibit higher sensitivity to noisy data, which is introduced by a considerable number of irrelevant documents in such scenarios.
\end{enumerate*}

\vspace{+3pt}
\noindent 
Our primary contributions can be summarized as follows:
\begin{itemize}[noitemsep,leftmargin=10px,nosep]
    \item \textbf{Exploration of fine-grained levels of support in citation evaluation:} to the best of our knowledge, we are the first to systematically investigate the effect of three-category support levels on faithfulness metrics in the task of automated citation evaluation. 
    \item  \textbf{Introduction of a comparative evaluation framework:} we propose a comparative evaluation framework designed to assess the alignment between metric scores and human judgments. This framework includes correlation analysis, classification, and retrieval evaluation to comprehensively evaluate the metric performance.
    \item  \textbf{Comprehensive experimental findings:} our experimental results demonstrate the best-performing faithfulness metrics still struggle to identify partially supporting citations, underscoring the inherent challenges of automated citation evaluation. Based on these findings, we offer practical recommendations for the development of more effective metrics.
\end{itemize}

\begin{figure*}[t]
    \centering
    \includegraphics[width=\linewidth]{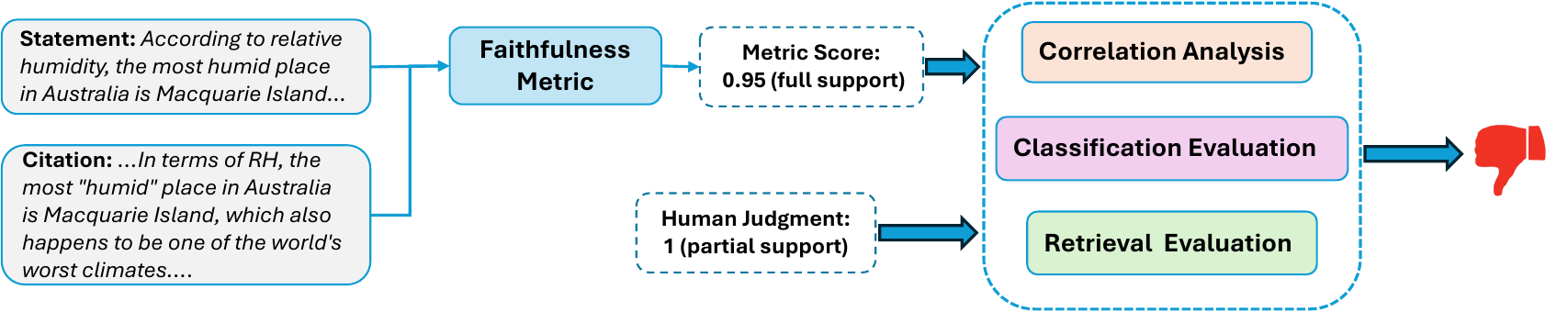}
    \caption{Our proposed comparative evaluation framework. A faithfulness metric assigns scores to given statements and their corresponding citations. Subsequently, our framework comprehensively assesses the alignment between these metric scores and human judgments by employing correlation analysis, classification, and retrieval evaluation.}
    \label{fig:framework}
\end{figure*}

\section{Evaluation Framework}

In this section, we introduce the proposed comparative evaluation framework. We begin by formalizing the task of automated citation evaluation. Subsequently, we detail three distinct evaluation protocols within this framework, ensuring a comprehensive assessment in alignment between faithfulness metrics and human judgments. Our framework is demonstrated in \autoref{fig:framework}.

\subsection{Task Formulation} 

The objective of automated citation evaluation is to automatically quantify the support level of a citation based on the citation and its associated statement.
In this work, we assume access to a dataset for automated citation evaluation, comprising pairs of statements and their corresponding citations, denoted as $(s_i,c_i)$. Each $s_i$ is a statement from the set $S$ of all statements produced by an LLM and each $c_i$ is a citation from a set $C$ of citations returned by the LLM.
According to human evaluation in the dataset, we categorize the citations into three distinct levels of support: full, partial, and no support. We adopt the definition of these levels of support from \citet{liu-etal-2023-evaluating}:
\begin{itemize}[noitemsep,leftmargin=10px,nosep]
    \item Full Support (\textsc{FS}): The citation fully supports every detail in the statement.
    \item Partial Support (\textsc{PS}): The citation supports certain aspects of the statement, while other details remain unsupported or are contradicted.
    \item No Support (\textsc{NS}):  None of the content in the statement is supported by the citation. For instance, the citation is entirely irrelevant or contradicts the statement.
\end{itemize}
\noindent
To this end, without loss of generality, we define a faithfulness metric as a scoring function, denoted as $F(s_i,c_i) \rightarrow \mathbb{R^+}$. 
For any given statement $s_i$ and its associated citation $c_i$, this scoring function provides a numeric score that indicates the extent of support provided by the citation to the statement.

\subsection{Evaluation Protocols}

The objective of evaluation protocols is to comprehensively assess the extent to which metric scores align with human judgments. In this work, we assess this alignment across three distinct dimensions: \textbf{correlation analysis}, \textbf{classification performance}, and \textbf{retrieval effectiveness}.

\subsubsection{Correlation Analysis} 

The correlation analysis aims to measure the general trend in the relationship between metric scores and human judgments.
Previous research~\cite{kryscinski-etal-2020-evaluating,pagnoni-etal-2021-understanding} has employed correlation analysis to meta-evaluate faithfulness metrics in abstractive text summarization. 
They involve measuring the extent to which metric scores align with binary levels of faithfulness, which are annotated by human assessors as either faithful ($1$) or unfaithful ($0$). 
Following them, we adapt correlation analysis to the task of automated citation evaluation. 
Specifically, Given statements and their associated citations, we assess how well predicted metric scores correlate with human-annotated support levels. 
To facilitate correlation analysis, we arbitrarily assign support levels $\{\textsc{FS}, \textsc{PS}, \textsc{NS}\}$ to values $\{0, 1, 2\}$. We then utilize standard correlation metrics such as the Pearson correlation coefficient to assess metric effectiveness. 

\subsubsection{Classification Evaluation} 

In addition to correlation analysis, we perform classification evaluation to determine the effectiveness of faithfulness metrics in discriminating citations based on their support level. 
Specifically, the metrics need to categorize a citation into one of three support levels: \textsc{FS}, \textsc{PS}, \textsc{NS}. 
Notably, existing faithfulness metrics do not apply to this three-way classification scenario, as they are unable to accurately determine the extent to which a statement is partially supported by its corresponding citation~\cite{honovich-etal-2022-true-evaluating}. 
To address this issue, we adopt a one-vs-one strategy, by effectively decomposing the three-way classification into three binary classification task settings:
\begin{enumerate*}[label=(\roman*)]
    \item Full Support vs.\ No Support (\isfullno{}),
    \item Full Support vs.\ Partial Support (\ispartialfull{}), and
    \item Partial Support vs.\ No Support (\ispartialno{}).
\end{enumerate*}
For each binary classification task setting, we construct a specialized dataset comprising only instances with corresponding binary support levels derived from the original dataset. 
We assess the performance of metrics on these tailored binary datasets using standard binary classification evaluation metrics such as ROC-AUC. 
The overall metric performance is then computed by averaging the results across all binary tasks.

\subsubsection{Retrieval Evaluation} 

The objective of retrieval evaluation is to measure the effectiveness of metrics in ranking citations according to their support levels.
This evaluation is motivated by the observation that previous correlation and classification evaluations presuppose the presence of citations within generated statements. 
However, real-world scenarios frequently present instances where citations are absent or irrelevant, highlighting the need for post-hoc retrieval to enhance citation quality~\cite{liu-etal-2023-evaluating,huang2023citation,huang2024traincitation}.
In post-hoc retrieval, candidate documents are retrieved to form a pool of potential citations using information retrieval techniques~\cite{karpukhin-etal-2020-dense}. 
Faithfulness metrics are then employed to rank citations based on their predicted metric scores, aiming to identify the citation with the highest support level~\cite{gao-etal-2023-rarr,bohnet2022attributed,gao-etal-2023-enabling}.
Ideally, a faithfulness metric should rank fully supporting citations at the top, followed by partially supporting citations, and finally non-supporting citations.
Similar to correlation analysis, we arbitrarily assign support levels $\{\textsc{FS}, \textsc{PS}, \textsc{NS}\}$ to relevance labels $\{2, 1, 0\}$. 
The metric effectiveness is assessed using standard information retrieval evaluation metrics, such as nDCG. This evaluation also provides a deeper understanding of metric performance in post-hoc citation retrieval scenarios.
\section{Faithfulness Metrics} 

In our experiments, we evaluate seven widely used faithfulness evaluation metrics, dividing them into similarity-based and entailment-based categories. Similarity-based metrics assess the level of support of a citation by measuring the degree of similarity between the citation and the associated statement. In contrast, entailment-based metrics leverage natural language inference (NLI) models~\cite{williams-etal-2018-broad} to estimate the support level based on the likelihood that the citation entails the statement.

\subsection{Similarity-Based Metrics}

\smallskip\noindent%
\textbf{\bertscore{}}
\cite{zhang2020bertscore} adopts BERT~\cite{devlin2019bert} to measure semantic similarity between a pair of text by aggregating cosine similarity among token-level BERT representation without further fine-tuning. 
We report the precision version of \bertscore since it correlates more with human judgments in faithfulness evaluation~\cite{pagnoni-etal-2021-understanding},  We use recommended \texttt{deberta-xlarge-mnli} \cite{he2021deberta} as the backbone of \bertscore{}.

\smallskip\noindent%
\textbf{\bartscore{}}
\cite{yuan2021bartscore} adopts BART~\cite{lewis2020bart} to measure the similarity between two texts based on conditional log-likelihood of generating target text from source text.
In our experiments, we leverage the faithfulness version of \bartscore, in which we treat the citation and the statement as the source and target text, respectively. We use the BART model fine-tuned on the CNN/DailyMail dataset~\cite{hermann2015cnndm} as the backbone of \bartscore{}.

\subsection{Entailment-Based Metrics}

\smallskip\noindent%
\textbf{\factcc{}} 
\cite{kryscinski-etal-2020-evaluating} is a BERT-based model to verify whether a generated text is faithful to a source text, which is fine-tuned on synthetic training data which contains simulated examples with different factual errors~\cite{kryscinski-etal-2020-evaluating}. This metric is also widely used for faithfulness evaluation in abstractive text summarization.

\smallskip\noindent%
\textbf{\summac{}} 
~\cite{laban-etal-2022-summac} is a RoBERTa-based model \cite{liu2019roberta} that is fine-tuned on NLI datasets. In this metric, a source text and its generated text are split into sentences. Entailment scores for all source/generated sentence pairs are then computed. Finally, the metric aggregates the scores of all pairs to obtain the final faithfulness score. 
The metric has two variants: 
\begin{enumerate*}[label=(\roman*)]
    \item \summaczsfull{} is a zero-shot version that is only pre-trained on NLI datasets; and
    \item \summacconvfull{} adds extra convolutional layers and is further fine-tuned on synthetic training data proposed in~\citet{kryscinski-etal-2020-evaluating}. We include both variants in our experiments.
\end{enumerate*}

\smallskip\noindent%
\textbf{\autoais{}}~\cite{honovich-etal-2022-true-evaluating,gao-etal-2023-rarr}
is a T5-11B~\cite{raffel2020exploring} model trained on a collection of NLI datasets, which is commonly used in recent automated citation evaluation~\cite{bohnet2022attributed,gao-etal-2023-rarr}. As the original output of \autoais{} is a numeric, either ``1'' (faithful) or ``0'' (unfaithful), we use the generated token probability of ``1'' as the predicted metric score.

\smallskip\noindent%
\textbf{\alignscore{}}~\cite{zha-etal-2023-alignscore} further fine-tunes a RoBERTa-based model~\cite{liu2019roberta} with a unified alignment loss function. To this end, a unified dataset containing a variety of related natural language processing datasets, such as NLI, question answering, and fact verification datasets, have been collected. In this work, we adapt the \texttt{large} version as it demonstrates the best performance.

\section{Experiments}
\label{sec:experiments}

In this section, we provide a description of the dataset statistics and the data processing method.
Subsequently, we discuss the evaluation metrics incorporated within our proposed framework, which assess the performance of faithfulness metrics in alignment with human judgments.

\subsection{Datasets}
\label{subsec:setup}

\begin{table}[tbp]
\caption{Data statistics of the \gense{} dataset. The dataset comprises 12,681 statement-citation pairs. Each pair has been annotated by human assessors based on three categories: full, partial, and no support.}
\label{tab:data_stats}
\centering
\footnotesize
\setlength\tabcolsep{5pt}
\centering
\small
\begin{tabular}{@{}lc@{}}
\toprule[1pt]
\textbf{Human Judgment} & \textbf{\# Statement-Citation Pair}  \\
\midrule
Full Support        & 6,616         \\
Partial Support      & 1,445           \\
No Support         & 4,620           \\
\midrule
Total         & 12,681     \\
\bottomrule[1pt]
\end{tabular}
\end{table}

\subsubsection{Data Statistics}

In the experiments, we employ the dataset of verifiability judgments~\cite{liu-etal-2023-evaluating} as our evaluation benchmark, referred to as \gense{}.
This dataset comprises a total of $12,681$ statement-citation pairs. For each pair, human assessors categorize the citation into one of three categories of support levels: full, partial, or no support. 
These categories indicate whether a citation provides full, partial, or no support to the associated statement.
The data statistics are illustrated in \autoref{tab:data_stats}.
Notably, for citations classified under the full or partial support categories, human assessors additionally extract explicit evidence from the citation that substantiates the associated statement.

\subsubsection{Data Processing}

While the \gense{} dataset aligns well with our research objectives, we encounter a significant challenge: the extensive length of most citations within the dataset. 
These citations often comprise a web document with thousands of words, far exceeding the maximum input capacity of most faithfulness metrics, which is limited to $512$ words.
This limitation necessitates input truncation, potentially compromising the reliability of faithfulness metrics.
To mitigate this issue, we adopt a strategy similar to previous studies~\cite{zha-etal-2023-alignscore}. 
Specifically, we segment each cited document into shorter text chunks, with a maximum length of $150$ words per chunk. These text chunks, along with their corresponding statements, serve as the inputs for faithfulness metrics to predicted metric scores.
Furthermore, to construct human judgments on the text chunks, we employ the Jaccard similarity index to identify text chunks containing human-annotated evidence, classifying them as either fully or partially supporting text chunks.

\subsection{Meta-Evaluation} 

For correlation evaluation, we report partial Pearson, Spearman, and Kendall coefficients, as recommended by previous research~\cite{pagnoni-etal-2021-understanding}.
In terms of classification evaluation, following previous studies~\cite{honovich-etal-2022-true-evaluating,ma-etal-2023-bump}, we report the Receiver Operating Characteristic-Area Under Curve (ROC-AUC) score as it obviates the need for manual threshold setting for each binary classification task.
Moreover, to capture the comprehensive performance across all binary classification tasks, we compute and report the macro-averaged ROC-AUC score.
For retrieval evaluation, we report standard information retrieval metrics: mean reciprocal rank (MRR) and normalized discounted cumulative gain (nDCG@n) scores.
\section{Results \& Analyses}
\label{subsec:results}

In this section, we discuss the results of the performance of faithfulness metrics across three distinct evaluation protocols. Following this, we integrate the observations derived from these evaluation protocols to discuss our main implications, offering practical recommendations to enhance metric effectiveness in automated citation evaluation.

\begin{table}[tbp]
\centering
\caption{Partial correlation coefficients between human-annotated levels of support and faithfulness metric scores on the \gense{} dataset. The best correlations are marked in bold.}
\begin{adjustbox}{max width=1.0\textwidth} {
\begin{tabular}{@{}lccc@{}}
\toprule[1pt]
\textbf{\textsc{Metric}}  & \textbf{Pearson} & \textbf{Spearman} & \textbf{Kendall}  \\
\midrule
\factcc{}           & 0.108 & -0.018 & -0.008   \\
\summaczsfull{}     & 0.326 & 0.143 & 0.106   \\
\bertscore{}        & 0.512 & 0.218 & 0.165   \\
\alignscore{}       & 0.551 & 0.275 & 0.199   \\
\bartscore{}        & 0.593 & 0.279 & 0.211 \\
\autoais{}          & \textbf{0.604} & 0.407 & 0.297   \\
\summacconvfull{}   & 0.565 & \textbf{0.444} & \textbf{0.342}  \\
\bottomrule[1pt]
\end{tabular}
}
\end{adjustbox}
\label{tab:correlation_results}
\end{table}

\subsection{Correlation Results} 

The correlation analysis results are demonstrated in \autoref{tab:correlation_results}. The following observations can be made:
\begin{enumerate*}[label=\arabic*)]
    \item the best-performing metrics reveal moderate correlations when analyzed using the Pearson coefficient. Specifically, \autoais{} achieves the highest Pearson coefficient, recording a value of $0.604$, marginally surpassing the second-best \bartscore{}, which posts a coefficient of $0.593$;
    \item there is a noticeable variation in correlation trends among high-performing metrics. Notably, \autoais{} shows a more substantial Pearson correlation, whereas \summacconvfull{} outperforms in Spearman and Kendall correlations. This divergence might be attributed to the Pearson coefficient assuming linear relationships between two variables. Such an assumption is often invalid in automated citation evaluation, rendering the Pearson coefficient less suitable for capturing the true relationships between metric scores and human judgments; and
    \item Generally, most metrics display relatively low Spearman and Kendall correlations compared to their Pearson correlations. For instance, \summacconvfull{} achieves the highest Spearman and Kendall correlations, with values of $0.445$ and $0.297$ respectively, which are considerably lower than its Pearson correlation of $0.565$. This disparity indicates that the metric scores of the best-performing metrics do not correlate well with human judgments, highlighting the limitations of existing metrics in scenarios involving fine-grained levels of support.
\end{enumerate*}

\begin{table}[tbp]
\caption{Classification performance of faithfulness metrics regarding ROC-AUC score (\%) on the \gense{} dataset. The overall performance is the macro-averaged performance of three binary classification settings. The best scores are marked in bold.}
\resizebox{0.985\linewidth}{!}{
\centering
\begin{tabular}{@{}lcccc@{}}
\toprule[1pt]
\textbf{\textsc{Metric}}  & \textbf{\isfullno{}} & \textbf{\ispartialfull{}} & \textbf{\ispartialno{}} & \textbf{\textsc{Overall}}  \\
\midrule
\factcc{}              & 68.77 & 62.58 & 56.81 & 62.72 \\
\summaczsfull{}     & 78.37 & 72.96 & 58.12 & 69.82 \\
\summacconvfull{}   & 85.32 & 78.74 & 62.57 & 75.54 \\ 
\bartscore{}           & 87.65 & 75.42 & 71.94 & 78.34 \\
\alignscore{}       & 90.97 & 81.41 & 70.53 & 80.97 \\
\bertscore{}           & 91.92 & 75.94 & \textbf{79.89} & 82.58 \\
\autoais{}          & \textbf{92.65} & \textbf{82.31} & 74.21 & \textbf{83.06}  \\ 
\bottomrule[1pt]
\end{tabular}
}
\label{tab:classification_results}
\end{table}

\subsection{Classification Results}

\autoref{tab:classification_results} presents the results of the classification evaluation.  The observations can be summarized as follows:
\begin{enumerate*}[label=\arabic*)]
    \item among all three binary classification task settings, most faithfulness metrics demonstrate superior performance in the \isfullno{} setting. Notably, entailment-based \autoais{}, with the highest ROC-AUC score of $92.65$, exemplifies significant discriminability between full support and no support instances. This performance can be attributed to its extensive parameters, comprising 11 billion parameters, in contrast to the hundreds of millions of other metrics;
    \item a pronounced decline in classification performance is observed across the other two settings. For instance, when comparing the \isfullno{} and \ispartialno{} settings, the ROC-AUC score of \autoais{} diminishes from $92.65$ to $74.21$. This decline indicates that even the best-performing metric struggles with granular sensitivity to varying levels of support; and
    \item while entailment-based \autoais{} generally surpasses other metrics in overall performance, it is outperformed by similarity-based \bertscore{} in the \ispartialno{} setting. Interestingly, while most metrics exhibit their lowest performance in this particular setting, \bertscore{} shows its least effectiveness in another setting, \ispartialfull{}. This underscores the unique prediction behaviors displayed by different types of metrics across the binary classification settings.
\end{enumerate*}

\subsection{Retrieval Results}

\autoref{tab:retrieval_results} presents the results of the retrieval evaluation. The key findings are as follows:
\begin{enumerate*}[label=\arabic*)]
    \item similarity-based metrics, \bartscore{} and \bertscore{}, outperforms other entailment-based metrics in both MRR and nDCG@n. For instance, entailment-based \autoais{} exhibits weaker MMR scores than \bartscore{} ($0.846$ vs. $0.881$). This is likely because entailment-based metrics are more sensitive to noisy information than similarity-based metrics, as many irrelevant documents exist in retrieval scenarios. This suggests the need for the robustness improvements of metrics in post-hoc retrieval scenarios;
    \item a significant correlation is observed between MRR and nDCG@n scores across all metrics. Notably, nDCG@n effectively captures the performance variations as the number of text chunks increases (i.e., 5, 10, 20). For instance, as the chunk count increases, \bartscore{} shows a marginal performance improvement (from $0.878$ to $0.897$), while \factcc{}—the least performing metric—exhibits a more pronounced enhancement (from $0.648$ to $0.710$); and
    \item we observe an intriguing shift in performance ranking when comparing metrics between classification and retrieval evaluations. For instance, \bartscore{} ascends from the fourth place to the first while \autoais{} sees a decline from the top to the third. This divergence highlights that these evaluations offer unique insights into the capabilities of metrics.
\end{enumerate*}

\begin{table}[t]
\centering
\caption{Retrieval performance of faithfulness metrics regarding MRR and nDCG@n scores on the \gense{} dataset. Note that we assign relevance labels $2$, $1$, and $0$ to full, partial, and no support, respectively. The best scores are marked in bold.}
\resizebox{0.985\linewidth}{!}{
\begin{tabular}{@{}lcccc@{}}
\toprule[1pt]
\textbf{\textsc{Metric}} & \textbf{MRR} & \textbf{nDCG@5} & \textbf{nDCG@10} & \textbf{nDCG@20} \\
\midrule
\factcc{}              & 0.656 & 0.648 & 0.689 & 0.710  \\
\summaczsfull{}     & 0.737 & 0.729 & 0.759 & 0.776  \\
\summacconvfull{}   & 0.776 & 0.772 & 0.798 & 0.811  \\
\alignscore{}       & 0.847 & 0.842 & 0.863 & 0.869  \\
\autoais{}          & 0.846 & 0.846 & 0.865 & 0.872  \\
\bertscore{}           & 0.867 & 0.865 & 0.881 & 0.887  \\
\bartscore{}           & \textbf{0.881} & \textbf{0.878} & \textbf{0.891} & \textbf{0.897}  \\
\bottomrule[1pt]
\end{tabular}
} 
\label{tab:retrieval_results}
\end{table}

\subsection{Implications}

Based on the evaluation results, we observe that no single faithfulness metric consistently excels across three distinct evaluation protocols. 
For instance, \summacconvfull{} achieves the highest performance in correlation analysis, yet it under-performs in classification and retrieval evaluations. 
This disparity suggests that these evaluation protocols are complementary and should be integrated to comprehensively assess the effectiveness of different metrics.
This inconsistency also reveals that the best-performing metrics are insufficiently effective in addressing fine-grained support level scenarios. Particularly, they fail to effectively distinguish partial support from either full or no support scenarios.
Furthermore, when comparing entailment-based metrics with similarity-based metrics, a notable shift in performance ranking is observed between the classification and retrieval evaluations. 
Specifically, the similarity-based \bartscore{} advances from fourth to first place, whereas the entailment-based \autoais{} declines from the top position to third place. 
This shift may be attributed to the higher sensitivity of entailment-based metrics to noisy data, which is introduced by irrelevant documents in retrieval scenarios.
This suggests the need for improving the robustness of entailment-based metrics against irrelevant documents.

Consequently, we propose the following practical recommendations to develop more effective metrics for automated citation evaluation:
\begin{enumerate*}[label=\arabic*)]
    \item \textbf{Development of training resources:} motivated by the observation that the best-performing metrics still struggle with identifying partial support, we recommend the development of training resources that include fine-grained support level annotations. These resources could significantly enhance the metrics' fine-grained sensitivity to varying support levels.
    \item \textbf{Introduction of contrastive learning:} to improve the robustness of metrics in post-hoc retrieval scenarios, we recommend fine-tuning metrics using contrastive learning frameworks. This method has demonstrated effectiveness in various information retrieval tasks~\cite{izacard2022contriever}.
\end{enumerate*}

\section{Related Work}

This section outlines two lines of related research: faithfulness evaluation metrics and citation evaluation.

\subsection{Faithfulness Evaluation Metrics}

Faithfulness evaluation metrics are crucial for assessing the factual consistency of text generated by models relative to the source text.
It receives great interest within the field of natural language generation (NLG)~\cite{huang2019novel,huang2021deepopht,zhang2021scaling,zhang2023tackling,huang2024optimizing,huang2024novel,zhu2024enhancing}, particularly in abstractive summarization~\cite{maynez-etal-2020-faithfulness,kryscinski-etal-2020-evaluating,huang2020query,huang2021gpt2mvs,zhang2024beyond,zhang2024qfmts}.
In general, faithfulness metrics are categorized into three types: entailment-based, similarity-based, and QA-based metrics. 
Entailment-based metrics employ natural language inference (NLI) models to determine if the source text entails the generated text~\cite{falke-etal-2019-ranking,laban-etal-2022-summac,honovich-etal-2022-true-evaluating,zha-etal-2023-alignscore}. 
Similarity-based metrics, such as BERTScore~\cite{zhang2020bertscore} and BARTScore~\cite{yuan2021bartscore}, quantify text similarity and have demonstrated robust performance in faithfulness evaluation~\cite{pagnoni-etal-2021-understanding,honovich-etal-2022-true-evaluating}.
QA-based metrics utilize a combination of question generation and question answering to estimate faithfulness levels~\cite{durmus-etal-2020-feqa,wang-etal-2020-asking,scialom-etal-2021-questeval,fabbri-etal-2022-qafacteval}.
In this work, we exclude QA-based metrics from our work, following recent works suggesting the challenging limitations in these metrics~\cite{kamoi-etal-2023-shortcomings}. %
We focus on the extrinsic evaluation of faithfulness metrics against human judgments in scenarios requiring fine-grained citation support.

\subsection{Citation Evaluation}

Citation evaluation seeks to enhance the trustworthiness of retrieval-augmented LLMs by verifying the support provided by citations to the generated statements~\cite{rashkin2021attribution,yue-etal-2023-automatic,huang2023citation,huang2024traincitation,zhang2024towards}.
Given the labor-intensive nature of manual citation evaluation, there has been a shift towards automated approaches to reduce dependence on human evaluation.
Since the goals of automated citation evaluation align closely with faithfulness evaluation in NLG, faithfulness metrics are employed to verify whether a citation supports the corresponding statement~\cite{li2023towards,sun2023towards,ye-etal-2024-effective,li2024attrbench,shen2024citekit,huang-etal-2024-learning}. 
Despite their widespread usage, the effectiveness of these metrics in more practical fine-grained citation support scenarios, such as those involving partial support by citations, has not been adequately addressed. Questions remain about the metrics' capability to differentiate citations in these fine-grained scenarios.
This work addresses these gaps by examining the effectiveness of faithfulness metrics across three distinct levels of citation support: full, partial, and no support.

\section{Conclusion}
LLMs are susceptible to generating hallucinated content, motivating the research on the integration of retrieval augmentation mechanisms to enhance the verifiability of generated statements through in-line citations. 
However, evaluating how well these citations support the statements remains a major challenge due to the labor-intensive nature of manual citation evaluation. 
Consequently, faithfulness metrics have been adopted to automate this evaluation, primarily determining citation support in a binary classification scenario. 
This paper proposes a comparative evaluation framework to explore the efficacy of faithfulness metrics beyond the binary scenario by examining three levels of citation support: full, partial, and no support.
Our framework assesses the alignment between metric scores and human judgments across three evaluation protocols: correlation analysis, classification evaluation, and retrieval evaluation.
Experimental results reveal that no single metric consistently excels across all evaluation protocols, indicating the complexity of automated citation evaluation and the limitations of existing faithfulness metrics in identifying partial support scenarios. 
Based on these findings, we further provide practical suggestions for the development of more effective metrics in automated citation evaluation.

\bibliographystyle{ACM-Reference-Format}
\bibliography{main}

\clearpage

\end{document}